\documentclass[aps,prd,nofootinbib,reprint,superscriptaddress,longbibliography]{revtex4-1}
\usepackage{etoolbox}
\usepackage{dcolumn}
\usepackage{bm}
\usepackage{graphics}
\usepackage{xcolor}
\usepackage{dirtytalk}
\usepackage{graphicx} 
\usepackage{blindtext}
\usepackage{amsmath}
\usepackage{amsfonts}
\usepackage{amssymb}
\usepackage{bbold}
\usepackage{mathtools,halloweenmath}
\usepackage[%
colorlinks=true,
urlcolor=blue,
linkcolor=red,
citecolor=blue
]{hyperref}
\usepackage{natbib}
\begin{document}
\title{\bf Gauge interactions and the Galilean limit}
\vskip 1cm
\author{Ashis Saha}
\email{ashis.saha@bose.res.in}
\affiliation{Department of Astrophysics and High Energy Physics,\linebreak
	S.N.~Bose National Centre for Basic Sciences,\linebreak
	JD Block, Sector-III, Salt Lake, Kolkata 700106, India}
\author{Rabin Banerjee}
\email{rabin@bose.res.in}
\affiliation{Department of Astrophysics and High Energy Physics,\linebreak
	S.N.~Bose National Centre for Basic Sciences,\linebreak
	JD Block, Sector-III, Salt Lake, Kolkata 700106, India}
\author{Sunandan Gangopadhyay}
\email{sunandan.gangopadhyay@bose.res.in,\\
	 sunandan.gangopadhyay@gmail.com}
\affiliation{Department of Astrophysics and High Energy Physics,\linebreak
	S.N.~Bose National Centre for Basic Sciences,\linebreak
	JD Block, Sector-III, Salt Lake, Kolkata 700106, India}	
	\begin{abstract}
\noindent The gauge invariant minimal couplings for a class of relativistic free matter fields with global symmetry (related to usual charge conservation) have been obtained by incorporating an iterative Noether mechanism. Non-relativistic reduction of both matter and gauge sectors of the obtained interacting theory is then performed simultaneously which in turn yield a set of new effective actions which are invariant under the Galilean relativistic framework. To be precise, we show that one can obtain the Schr\"odinger field theory coupled to Galilean electromagnetism from the scalar quantum electrodynamics theory. Higher derivative corrections have also been included for which the non-relativistic reductions have been consistently carried out once again. On the other hand, the action for quantum electrodynamics leads to the Galilean Pauli-Schr\"odinger theory where the gauge field is non-relativistic or Galilean. Further, some novel relations are found (in both the electric and magnetic limits) between various components appearing in the Galilean avatar of electrodynamics.
\end{abstract}
\maketitle
\section{Introduction}
 The gauge principle, which requires the invariance of a physical theory under local gauge transformations is fundamental for the devolopment of modern quantum field theory and gravity, both relativistic and non-relativistic \cite{hooft1994under}. It forms the basis for introducing interactions in a field theory of matter invariant under global symmetry transformations by the prescription of replacing ordinary derivatives by covariant derivatives such that the extra correction terms balances the difference of the transformation of the fields at two nearby points. Now field strengths or curvatures are defined using the Ricci identity from which the action can be written. Thus, the origins of the gauge principles are more geometrical rather than physical. The introduction of interactions remain somewhat unclear and hence less physical.\\
 A more physical approach was presented by Deser \cite{Deser:1969wk} where the role of interactions, rather than the gauge principle is crucial. It is based on an iterative Noether prescription where interactions are introduced in a step by step method. The idea is to calculate the source for a particular step from the source of the earlier step. This iteration stops when no new source is obtained. This method therefore acts like a bridge bwteen the two theorems of Noether. Beginning from a theory that has a global symmetry (Noether's first theorem), the iteration finally yields an interacting theory invariant under local symmetry transformations (Noether's second theorem). There are many examples in the literature where this method has been applied which naturally gives the minimal coupling \cite{Deser:1979zb,Deser:1995kp,Khoudeir:2001yw,Khoudeir:1996mi,Basu:2017idu,Bhattacharjee:2013kja,Banerjee:2021lqe}. In this work, we consider a class of well-known relativistic matter actions which have global $U(1)$ phase symmetries, namely, the free complex ($\mathbb{C}$) scalar field action, $\mathbb{C}$-scalar field action with higher derivative corrections and the free Dirac field action. We then introduce the gauge interaction terms by following the approach advocated by Deser.\\
 On the other hand, non-relativistic reduction of Poincare invariant field actions in turn gifts us an effective low-energy description. In the context of diffeomorphism invariant theories, this reduction has always been a matter of great interest \cite{Andreev:2013qsa,Andreev:2014gia,Jensen:2014wha,Banerjee:2014pya,Banerjee:2015tga,Banerjee:2015rca,Banerjee:2016bbm}. However, because of the appearance of a universal time, it is challenging to come up with a systematic formulation for the derivation. A possible way is to follow Dirac's procedure to obtain the Galilean invariant wave equation \cite{Levy-Leblond:1967eic}. Alternatively, creating a methodical algorithm to construct the NR limit of the corresponding relativistic theory is also well-known. In general, this reduction procedure is mainly algebraic rather than kinematic. For instance, using the $c\rightarrow\infty$ limit reduces the quantum mechanical Poincare generators to the Galilean generators. This contraction is known as the In\"on\"u-Wigner group contraction in which the number of generators for the two theories (relativistic and NR) remains same. On a similar note, another method is the Eisenhart-Duval lift in which the reduction is done by Kaluza-Klein type null reduction from one higher dimension \cite{Eisenhart,Duval:1984cj,Duval:1990hj,Cariglia:2016oft,Banerjee:2014nja}. Apart from the purely algebraic reductions, investigations for reduction at the kinematic level have also been done. Some works in this direction can be found in \cite{Banerjee:2018gqz,Sharma:2023chs}. In \cite{Banerjee:2018pvs,Banerjee:2019hec}, it was shown that starting from a relativistic matter field action one can systematically obtain the corresponding non-relativistic matter action which is consistent with the algebraic level reduction. In this work, we have incorporated this method in order to perform the non-relativistic reduction of the relativistic matter fields we have considered. A nice review on non-relativistic quantum field theories can be found in \cite{Baiguera:2023fus}.\\
 In recent times, studies related to electrodynamics which is invariant under Galilean transformations have gained a lot of attention. It was first addressed by Le Bellac and Levy-Leblond in the field theoretic set up \cite{LeBellac:1973unm} and it was later revisited in \cite{Santos:2004pq} where embedding techniques were used. Some interesting works related to this particular topic can be found in \cite{2013EPJP,Bagchi:2014ysa,VandenBleeken:2015rzu,Bergshoeff:2015sic,Festuccia:2016caf,Mehra:2021sfx,Chapman:2020vtn,Banerjee:2022uqj,Baiguera:2022cbp,Fontanella:2024hgv}. In \cite{Banerjee:2022eaj}, a systematic derivation of the electric and magnetic limit from the Lorentz transformation of an arbitrary four vector was given. Further, it was done for both contravariant and covariant component of the four vector which in turn led to a distinct set of non-relativistic maps for both electric and magnetic limits. This study was further extended in \cite{Banerjee:2023koh,Banerjee:2023wcp}. We have used these non-relativistic maps for Maxwell electrodynamics to perform the non-relativistic reduction of the gauge sector of the obtained interacting theory.\\
 The organization of this paper is as follows. In Sec.\eqref{Sec1}, starting from a relativistic $\mathbb{C}$-scalar field theory, the gauge interactions have been introduced by using Deser's method. Consideration of non-relativistic reduction of both matter and gauge sector of the obtained interacting theory then lead us to the Galilean version of it. The same treatment for the $\mathbb{C}$-scalar field theory with higher derivative corrections have been provided in Sec.\eqref{Sec2}. Finally, in Sec.\eqref{Sec3} we consider the free Dirac field theory and obtain the action for quantum electrodynamics from it. Further, non-relativistic reductions then lead us to the Galilean Pauli-Schr\"odinger theory. We conclude in Sec.\eqref{Sec4}. 
  \section{Complex-scalar field theory}\label{Sec1}
  \subsection{Gauge interactions from an iterative Noether approach}
  In \cite{Deser:1969wk}, it was shown that without effectively using the gauge principle, one can construct a gauge theory by incorporating an iterative Noether approach. The resulting theory is invariant under local gauge transformations. Let us understand the method with the help of an example. Consider a complex ($\mathbb{C}$) scalar field theory characterized by the following action
  \begin{eqnarray}\label{complexaction}
  S=\int c~dt\int d^3x~\left(\partial_{\mu}\phi\partial^{\mu}\phi^{\star}-m^2\phi^{\star}\phi\right)~.
  \end{eqnarray}
  The above action has global $U(1)$ symmetry, more precisely, it is invariant under the following transformations 
  \begin{eqnarray}
  \phi\rightarrow e^{i\eta}\phi;~\phi^{\star}\rightarrow e^{-i\eta}\phi^{\star}~.
  \end{eqnarray}
  This symmetry in turn yields the conservation law $\partial_{\mu}j^{\mu}=0$, which is nothing but the manifestation of Noether's first theorem. Now, the standard way to construct a theory which will be invariant under local $U(1)$ symmetry is done by considering the transformation parameter $\eta$ local $\eta(x^{\mu})$
  and then incorporating the gauge principle to replace the usual derivatives by covariant derivatives. However, the same result can also be obtained from the iterative Noether approach. To do this, we first write down space and time parts of the largrangian density $\mathcal{L}_0$ of the theory (given in the action \eqref{complexaction}) separately. This leads
  \begin{eqnarray}\label{Mlag}
  	\mathcal{L}_0 = \mathcal{L}_t+\mathcal{L}_s \equiv \partial_0\phi~\partial_0\phi^{\star}-\partial_i\phi~\partial_i\phi^{\star}-m^2\phi^{\star}\phi
  \end{eqnarray}
  where $\mathcal{L}_t$ represents the time part of the lagrangian density $\mathcal{L}_t=\partial_0\phi~\partial_0\phi^{\star}$ and the space part is being represented by $\mathcal{L}_s=-\partial_i\phi~\partial_i\phi^{\star}-m^2\phi^{\star}\phi$. We would also like to mention that we are using following signature for the background flat metric $(1,-1,-1,-1)$. According to Noether's first theorem, the currents associated to a global symmetry are
  \begin{eqnarray}
  \eta j^0&=&\frac{\partial\mathcal{L}_t}{\partial(\partial_0\phi)}\delta\phi+c.c.\label{Noether1}\\
  \eta j^i&=&\frac{\partial\mathcal{L}_t}{\partial(\partial_i\phi)}\delta\phi+c.c.~.\label{Noether2}
  \end{eqnarray}
 By using the above formulae, it is pretty straight forward to show the following result for the time part
 \begin{eqnarray}
 j^0=j_0=i\left(\phi\partial_0 \phi^{\star}-\phi^{\star}\partial_0\phi\right)~.
 \end{eqnarray}
 With the above result in hand, the first correction to the time part is
 \begin{eqnarray}
 \mathcal{L}^{(1)}_t=j^0\mathcal{A}_0=i\left(\phi\partial_0 \phi^{\star}-\phi^{\star}\partial_0\phi\right)\mathcal{A}_0~.
 \end{eqnarray}
 By substituting $\mathcal{L}_t$ by $\mathcal{L}^{(1)}_t$ in eq.\eqref{Noether1}, we once gain get the local current corresponding to the above term. This reads
 \begin{eqnarray}\label{1stcurrent}
 j^{(1)}_0= 2\phi^{\star}\phi \mathcal{A}_0~.
 \end{eqnarray}
 This in turn lead us to the next correction for the time part which has the following form
\begin{eqnarray}\label{Lt2}
 \mathcal{L}^{(2)}_t=\frac{1}{2}j^{(1)}_0 \mathcal{A}_0=\phi^{\star}\phi \mathcal{A}_0^2
 \end{eqnarray}
 where we have introduced a half factor in the above expression necessary to reproduce eq.\eqref{1stcurrent} from the variation of the term $\int \mathcal{L}^{(2)}_t$ with respect to $\mathcal{A}_0$, that is
 \begin{eqnarray}\label{fvar}
 j^{(1)}_0 = \frac{\delta}{\delta \mathcal{A}_0}\int \mathcal{L}^{(2)}_t~.
 \end{eqnarray}
 From the obtained expression of the second correction $\mathcal{L}^{(2)}_t$ (appearing in eq.\eqref{Lt2}), we can observe that if we use the formula for current (given in eq.\eqref{Noether1}), it will have a vanishing result which is due to the fact that it has no derivative of the fields. Collecting all the correction terms, one can now write
 \begin{eqnarray}\label{tcov}
 \mathcal{L}_{\mathrm{time}}&=&\mathcal{L}_t+\mathcal{L}^{(1)}_t+\mathcal{L}^{(2)}_t\nonumber\\
 &=&\partial_0\phi~\partial_0\phi^{\star}+i\left(\phi\partial_0 \phi^{\star}-\phi^{\star}\partial_0\phi\right)\mathcal{A}_0+\phi^{\star}\phi \mathcal{A}_0^2\nonumber\\
 &=&(\mathcal{D}_0\phi)(\mathcal{D}_0\phi)^{\star}~.
 \end{eqnarray}
 In the last line we have defined $\mathcal{D}_0=\partial_0+i\mathcal{A}_0$ in order to display the minimally coupled structure. Now let us move onto the space part of the lagrangian density $\mathcal{L}_s$ (given in eq.\eqref{Mlag}). In this case, the first correction term is obtained to be
 \begin{eqnarray}
 \mathcal{L}_s^{(1)}=j^i\mathcal{A}_i=-j_i\mathcal{A}_i=i\left(\phi^{\star}\partial_i\phi-\phi\partial_i\phi^{\star}\right)\mathcal{A}_i~.
 \end{eqnarray}
 Similar to the time part, we now substitute $\mathcal{L}_s$ by $\mathcal{L}_s^{(1)}$ in eq.\eqref{Noether2} and obtain the following expression of corresponding local current
 \begin{eqnarray}
 j_i^{(1)}=-2\phi^{\star}\phi \mathcal{A}_i~.
 \end{eqnarray}
 This in turn means our second correction term for the space part will be 
 \begin{eqnarray}
 \mathcal{L}_s^{(2)}=\frac{1}{2}  j_i^{(1)}\mathcal{A}_i=-\phi^{\star}\phi \mathcal{A}_i^2
 \end{eqnarray}
 where we have once again introduced a half factor to the definition of the interaction term due to the same logic we have explained in eq.\eqref{fvar}. As $\mathcal{L}_s^{(2)}$ does not have derivative, the iteration process stops here. Once again by collecting all the correction terms, we write the following
 \begin{eqnarray}\label{scov}
 \mathcal{L}_{\mathrm{space}}&=&\mathcal{L}_s+\mathcal{L}^{(1)}_s+\mathcal{L}^{(2)}_s\nonumber\\
 &=&-\partial_i\phi~\partial_i\phi^{\star}-m^2\phi^{\star}\phi+i\left(\phi^{\star}\partial_i\phi-\phi\partial_i\phi^{\star}\right)\mathcal{A}_i\nonumber\\
 &&-\phi^{\star}\phi \mathcal{A}_i^2\nonumber\\
 &=&-(\mathcal{D}_i\phi)(\mathcal{D}_i\phi)^{\star}-m^2\phi^{\star}\phi
 \end{eqnarray}
 where we have defined $\mathcal{D}_i=\partial_i+i\mathcal{A}_i$ in order to obtain the minimally coupled structure of the space part. Now, substituting the complete interacting lagrangian density (sum of eq.\eqref{tcov} and eq.\eqref{scov}) in the action (given in eq.\eqref{complexaction}) we obtain
  \begin{eqnarray}\label{gaugedaction}
 S_{\mathrm{matter}}=\int c~dt\int d^3x~\left[\left(\mathcal{D}_{\mu}\phi\right)\left(\mathcal{D}^{\mu}\phi\right)^{\star}-m^2\phi^{\star}\phi\right]~.
 \end{eqnarray}
 The above action is invariant under the following local gauge transformations
 \begin{eqnarray}
 	\phi\rightarrow e^{i\eta(x)}\phi,&&~\phi^{\star}\rightarrow e^{-i\eta(x)}\phi^{\star},\nonumber\\
 	\mathcal{A}_{\mu}\rightarrow &&\mathcal{A}_{\mu}-\partial_{\mu}\eta(x)~.
 \end{eqnarray}
 If we now include the dynamics of the gauge field $A_{\mu}$ which is represented by the traditional Maxwell term, this reads
 \begin{eqnarray}\label{MaxwellAction}
 S_{\mathrm{SQED}}&=&S_{\mathrm{matter}}+S_{\mathrm{gauge}}\nonumber\\
 &&=\int c~dt\nonumber\\
 &&\int d^3x\left[\left(\mathcal{D}_{\mu}\phi\right)\left(\mathcal{D}^{\mu}\phi\right)^{\star}-m^2\phi^{\star}\phi-\frac{1}{4}F_{\mu\nu}F^{\mu\nu}\right]~~~~~
 \end{eqnarray}
 where $F_{\mu\nu}=\partial_{\mu}\mathcal{A}_{\nu}-\partial_{\nu}\mathcal{A}_{\mu}$. The above action is commonly denoted as scalar quantum electrodynamics (SQED) action. Our next aim is to consider non-relativistic limits of both matter ($\mathbb{C}$-scalar field) and the gauge sector ($\mathcal{A}_{\mu}$) of the above obtained SQED theory.
 \subsection{Non-relativistic reduction of the theory}
 It is now a well-known fact that in the limit $c\rightarrow\infty$, a Lorentz invariant relativistic action reduces to a non-relativistic action which is invariant under Galilean relativistic transformations. First, we perform this reduction for the matter sector of the SQED action, given in eq.\eqref{gaugedaction}. It is to be noted that before we consider the limit $c\rightarrow\infty$, one needs to restore dimensionfull quantity $c$ in the mentioned action in the following way
 \begin{eqnarray}
 \partial_0=\frac{1}{c}\partial_t;~\mathcal{A}_{\mu}=(\mathcal{A}_0,\mathcal{A}_i)\equiv\left(\frac{A_0}{c},A_i\right)~.
 \end{eqnarray}
 In order to do this we make use of the following map to introduce the Schr\"odinger fields \cite{Banerjee:2018pvs,Banerjee:2019hec}
 \begin{eqnarray}\label{NRmap}
 \phi&=&\frac{1}{\sqrt{2mc}}e^{-imc^2t}\psi(t,x^i);\nonumber\\
 \phi^{\star}&=&\frac{1}{\sqrt{2mc}}e^{imc^2t}\psi^{\star}(t,x^i)~.
 \end{eqnarray}
This in turn implies the following results
\begin{eqnarray}\label{NRQuantity}
	\partial_t\phi&=&\frac{1}{\sqrt{2mc}}e^{-imc^2t}\left(\partial_t\psi(t,x^i)-imc^2\psi(t,x^i)\right)\nonumber\\
	\partial_i\phi&=&\frac{1}{\sqrt{2mc}}e^{-imc^2t}\partial_i\psi(t,x^i)\nonumber\\
	\partial_t\phi^{\star}&=&\frac{1}{\sqrt{2mc}}e^{imc^2t}\left(\partial_t\psi^{\star}(t,x^i)+imc^2\psi^{\star}(t,x^i)\right)\nonumber\\
	\partial_i\phi^{\star}&=&\frac{1}{\sqrt{2mc}}e^{imc^2t}\partial_i\psi^{\star}(t,x^i)~.
\end{eqnarray}
\begin{widetext}
We now substitute the above results in the action $S_{\mathrm{matter}}$ (given in \eqref{gaugedaction}) and obtain the following form
\begin{eqnarray}
S_{\mathrm{matter}}&=&\int dt\int d^3x\left(\frac{1}{2m}\right)\Bigg[\frac{1}{c^2}(\partial_t\psi)(\partial_t\psi^{\star})+im\psi^{\star}\overleftrightarrow{\partial_t}\psi+(mc)^2\psi^{\star}\psi-i\left(\frac{A_0}{c^2}\psi\overleftrightarrow{\partial_t}\psi^{\star}\right)\nonumber\\
&&-2A_0m\psi^{\star}\psi-\left(\partial_i\psi\right)\left(\partial_i\psi^{\star}\right)+iA_i\psi^{\star}\overleftrightarrow{\partial}_i\psi-(mc)^2\psi^{\star}\psi+A_i^2\psi^{\star}\psi+\left(\frac{A_i}{c}\right)^2\psi^{\star}\psi\Bigg]
\end{eqnarray}
where $\psi^{\star}\overleftrightarrow{\partial}_i\psi=\psi^{\star}\left(\partial_i\psi\right)-\psi\left(\partial_i\psi\right)^{\star}$.
We now take the $c\rightarrow\infty$ limit of the above action and obtain the following reduced form
\begin{eqnarray}\label{gaugedGalilean}
S_{\mathrm{matter}}^{\mathrm{NR}}=\int dt\int d^3x \left(\frac{i}{2}\right)\left[\psi^{\star}\overleftrightarrow{D_t}\psi+\frac{i}{m}(D_i\psi)(D_i\psi)^{\star}\right]\nonumber\\
\end{eqnarray}
\end{widetext}
where we have defined $D_t=\partial_t+iA_0$ and $D_i=\partial_i+iA_i$ in order to introduce minimally coupled form in the Galilean relativistic set up. Furthermore, in the above we have introduced the following
\begin{eqnarray}
	\psi^{\star}\overleftrightarrow{D_t}\psi=\psi^{\star}\left(D_t\psi\right)-\psi\left(D_t\psi\right)^{\star}~.
\end{eqnarray}
This action is usually denoted as the Galilean relativistic Schr\"odinger field action. The above obtained action matches perfectly with the one shown in \cite{Banerjee:2021lqe}. So far we have considered non-relativistic reduction of matter sector only and we observe that even in the Galilean limit of the $\mathbb{C}$-scalar field, it is still possible to define covariant derivatives and the dynamics of the gauge field is still given by the usual Maxwell term (given in eq.\eqref{MaxwellAction}). We now move on to consider non-relativistic limit of the gauge sector of both of the matter action $S_{\mathrm{matter}}^{\mathrm{NR}}$ and the dynamical Maxwell term, by following the approach shown in \cite{Banerjee:2022eaj}.
\subsubsection{Electric limit}
First, we make use of the following maps introduced in \cite{Banerjee:2022eaj}
\begin{eqnarray}\label{Elimit}
	&&A_0\rightarrow\frac{a_0}{c};~A_i\rightarrow a_i~(\mathrm{covariant}~\mathrm{maps})\nonumber\\
	&&A^0\rightarrow ca^0;~A^i\rightarrow a^i~(\mathrm{contravariant}~\mathrm{maps})~.
\end{eqnarray}
This is known as the electric limit of the gauge field $A_{\mu}$ as it produces largely timelike vectors. We substitute the above maps in the non-relativistic matter action obtained in eq.\eqref{gaugedGalilean} and consider the limit $c\rightarrow\infty$. This in turn yields the following fully non-relativistic action in the Galilean electric limit
\begin{eqnarray}\label{Ecovar}
S_{\mathrm{matter}}^{\mathrm{NR}}&=&\int dt\nonumber\\
&&\int d^3x \left(\frac{i}{2}\right)\left[\psi^{\star}\overleftrightarrow{\partial_t}\psi+\frac{i}{m}(\Lambda_i\psi)(\Lambda_i\psi)^{\star}\right]~~
\end{eqnarray}
where $\Lambda_i=\partial_i+ia_i$. This result is very interesting in the sense that the time-covariance is broken once we consider the Galilean reduction of the gauge field. However, space-covariance still survives. This is not unexpected as we know in Galilean relativity time and space do not share same footing, the absolute nature of time in Galilean relativity invokes invariance not covariance. Further, the dynamics of the gauge field is now being represented by the following modified Maxwell Lagrangian density \cite{Banerjee:2022eaj}
\begin{eqnarray}\label{MaxwellE}
\mathcal{L}_{\mathrm{EM}}^{\mathrm{NR}}= \frac{1}{2}\partial^ia^0\left(\partial_ta_i-\partial_ia_0\right)-\frac{1}{4}f_{ij}f^{ij}
\end{eqnarray}
where $f_{ij}=\partial_ia_j-\partial_ja_i$~.
\begin{widetext}
This in turn yields the final form of the SQED action (given in eq.\eqref{MaxwellAction}) in the full Galilean setup (in the electric limit)
\begin{eqnarray}\label{NRSQED}
S_{\mathrm{SQED}}^{\mathrm{NR}}&=&\int dt\int d^3x\left(\frac{i}{2}\right)\left[\psi^{\star}\overleftrightarrow{\partial_t}\psi+\frac{i}{m}(\Lambda_i\psi)(\Lambda_i\psi)^{\star}-i\partial^ia^0\left(\partial_ta_i-\partial_ia_0\right)+\frac{i}{2}f_{ij}f^{ij}\right]~.
\end{eqnarray}
\end{widetext}
It can be noted that in obtaining the above, we have only used covariant map of eq.\eqref{Elimit}. However, one can also use the contravariant maps for the same task. In order to check this we write down the matter action given in eq.\eqref{gaugedGalilean} in the following way
\begin{widetext}
\begin{eqnarray}\label{Econtra}
S_{\mathrm{matter}}^{\mathrm{NR}}&=&\int dt\int d^3x \left(\frac{i}{2}\right)\left[\psi^{\star}\left(D_t\psi\right)-\psi\left(D_t\psi\right)^{\star}+\frac{i}{m}(D_i\psi)(D_i\psi)^{\star}\right]\nonumber\\
&=&\int dt\int d^3x \left(\frac{i}{2}\right)\left[\psi^{\star}\left(\partial_t\psi\right)-\psi\left(\partial_t\psi\right)^{\star}+2i\eta_{0\rho}A^{\rho}\psi^{\star}\psi+\frac{i}{m}|\left(\partial_i+i\eta_{ij}A^{j}\right)\psi|^2\right]\nonumber\\
&=&\int dt\int d^3x \left(\frac{i}{2}\right)\left[\psi^{\star}\left(\partial_t\psi\right)-\psi\left(\partial_t\psi\right)^{\star}+2iA^{0}\psi^{\star}\psi+\frac{i}{m}|\left(\partial_i-iA^{i}\right)\psi|^2\right]\nonumber\\
&=&\int dt\int d^3x \left(\frac{i}{2}\right)\left[\psi^{\star}\left(\partial_t\psi\right)-\psi\left(\partial_t\psi\right)^{\star}+\left(2ica^{0}\right)\psi^{\star}\psi+\frac{i}{m}|\left(\partial_i-ia^i\right)\psi|^2\right]
\end{eqnarray}
where in the last line we have used the contravariant map of the electric limit (given in eq.\eqref{Elimit}).
\end{widetext}
It can be noted that if we now proceed to consider $c\rightarrow\infty$ limit, it apparently seems that the above action diverges. However, this cannot be true as it will then imply that the covariant map and the contravariant map lead to different results and if so then the consistency of the entire procedure break down. This discomfort can be removed only if there exists a relation between the non-relativistic components $a_0$ and $a^0$. One could derive this relation by demanding the fact that the results obtained by using the contravariant map has to match with the same obtained by using the covariant map. This in turn leads to the following relations 
\begin{eqnarray}\label{ElimitRelations}
a^0=\frac{1}{c^2}a_0;~a^i=-a_i~.
\end{eqnarray}
These relations can also be derived from a simpler origin by noticing that $A_0=\eta_{0\mu}A^{\mu}=\eta_{00}A^0=A^0$. Now, by substituting the maps for the contravariant and covariant components in the electric Galilean limit gives the first relation in eq.\eqref{ElimitRelations}. The second relation can also be derived in a similar way. A word of caution is in order now. It should be noted that using the map $a^0=\frac{1}{c^2}a_0$ in eq.\eqref{MaxwellE} gives a term $\frac{1}{c^2}\partial^ia_0\left(\partial_ta_i-\partial_ia_0\right)$ which may look like a sub-leading term. However, one should not take the $c\rightarrow\infty$ limit now and remove this term from eq.\eqref{MaxwellE} altogether. The $c\rightarrow\infty$ limit has already been taken when we moved from the relativistic components to non-relativistic components.
We dub the relations in eq.\eqref{ElimitRelations} as the electric limit relations and these can also be written as
\begin{eqnarray}
a^{\mu}=\mathcal{J}^{\mu\nu}a_{\nu}
\end{eqnarray}
where $\mathcal{J}^{\mu\nu}$ has the following form
\begin{eqnarray}
\mathcal{J}^{\mu\nu}=\begin{pmatrix}
\frac{1}{c^2} & 0 & 0& 0\\
0 & -1 & 0 & 0\\
0 & 0 & -1 & 0\\
0 & 0 & 0 & -1\\
\end{pmatrix}~.
\end{eqnarray}
We now use these relations in the action (obtained in \eqref{Econtra}) and after takng the $c\rightarrow\infty$ limit, one obtains the same matter action and non-relativistic SQED action that we have shown in eq.\eqref{Ecovar} and eq.\eqref{NRSQED} respectively.
\subsubsection{Magnetic limit}
In the magnetic limit, one uses the following maps \cite{Banerjee:2022eaj}
\begin{eqnarray}\label{Mlimit}
&&A_0\rightarrow-ca_0;~A_i\rightarrow a_i~(\mathrm{covariant}~\mathrm{maps})\nonumber\\
&&A^0\rightarrow -\frac{a^0}{c};~A^i\rightarrow a^i~(\mathrm{contravariant}~\mathrm{maps})~.
\end{eqnarray}
\begin{widetext}
We first use the contravariant maps in the matter action given in eq.\eqref{gaugedGalilean}. This in turn leads us to the following
\begin{eqnarray}\label{Mcontra}
S_{\mathrm{matter}}^{\mathrm{NR}}&=&\int dt\int d^3x \left(\frac{i}{2}\right)\left[\psi^{\star}\left(\partial_t\psi\right)-\psi\left(\partial_t\psi\right)^{\star}-\frac{2ia^{0}}{c}\psi^{\star}\psi+\frac{i}{m}|\left(\partial_i-ia^{i}\right)\psi|^2\right]\nonumber\\
&=&\int dt\int d^3x \left(\frac{i}{2}\right)\left[\psi^{\star}\overleftrightarrow{\partial_t}\psi+\frac{i}{m}(\tilde{\Lambda}_i\psi)(\tilde{\Lambda}_i\psi)^{\star}\right];~\tilde{\Lambda}_i=\partial_i-ia^i
\end{eqnarray}
where we have taken the limit $c\rightarrow\infty$ in the last line. The above action \textcolor{red}{is} quite similar to the one we have obtained from the electric limit but the form of the covariant derivative has changed. Next, we intend to use the covariant maps of the magnetic limit. This yields
\begin{eqnarray}\label{magneticcova}
S_{\mathrm{matter}}^{\mathrm{NR}}&=&\int dt\int d^3x \left(\frac{i}{2}\right)\left[\psi^{\star}\left(\partial_t\psi\right)-\psi\left(\partial_t\psi\right)^{\star}-2a_0c\psi^{\star}\psi+\frac{i}{m}|\left(\partial_i+ia_{i}\right)\psi|^2\right]~.
\end{eqnarray}
Once again, it can be noted that in the limit $c\rightarrow\infty$, the above action diverges. In an Appendix, we derive the interacting Schr\"odinger theory from the non-relativistic free theory using Deser's iterative Noether approach.
\end{widetext}
 We now follow the same argument that we have already shown for the case of electric limit and this in turn leads to the following relations in context of magnetic limit
\begin{eqnarray}\label{MlimitRelations}
a_0=\frac{1}{c^2}a^0;~a_i=-a^i~.
\end{eqnarray}
Similar to the electric limit, the above relations have a simple origin as can be seen from the fact $A^0=\eta^{00} A_0=A_0$. By using the contravariant and covariant maps one obtains $a_0=\frac{1}{c^2}a^0$. We dub these relations as the magnetic limit relations which can be written as
\begin{eqnarray}
a_{\mu}=\mathcal{G}_{\mu\nu}a^{\nu}
\end{eqnarray}
where $\mathcal{G}_{\mu\nu}$ has the following form
\begin{eqnarray}
\mathcal{G}_{\mu\nu}=\begin{pmatrix}
\frac{1}{c^2} & 0 & 0& 0\\
0 & -1 & 0 & 0\\
0 & 0 & -1 & 0\\
0 & 0 & 0 & -1\\
\end{pmatrix}~.
\end{eqnarray}
It can observed that the relation between the spatial components are similar to that we have obtained for the same in the electric limit. However, the relation between the time-components are different from the one we got in context of the electric limit. The above relations lead to the same result which has been obtained from the contravariant maps (given in eq.\eqref{Mcontra}). Now, in the magnetic limit, the dynamics of the gauge field is being represented by the following form of the Maxwell lagrangian density \cite{Banerjee:2022eaj}
\begin{eqnarray}\label{MaxwellM}
\mathcal{L}_{\mathrm{EM}}^{\mathrm{NR}}= \frac{1}{2}\partial_ia_0\left(\partial_ta^i-\partial^ia^0\right)-\frac{1}{4}f_{ij}f^{ij}~.
\end{eqnarray}
We now write down the form of the action corresponding to the non-relativistic SQED theory (in the magnetic limit)
\begin{widetext}
\begin{eqnarray}
S_{\mathrm{SQED}}^{\mathrm{NR}}&=&\int dt\int d^3x\left(\frac{i}{2}\right)\left[\psi^{\star}\overleftrightarrow{\partial_t}\psi+\frac{i}{m}(\tilde{\Lambda}_i\psi)(\tilde{\Lambda}_i\psi)^{\star}-i\partial_ia_0\left(\partial_ta^i-\partial^ia^0\right)+\frac{i}{2}f_{ij}f^{ij}\right]~.
\end{eqnarray}
\end{widetext}
\section{Complex scalar field theory with higher derivatives}\label{Sec2}
We now proceed to consider a system with more involved structure, that is, a $\mathbb{C}$-scalar field theory with higher derivative terms.
\begin{widetext}
\subsection{Gauge interactions from an iterative Noether approach}
We now follow the same footsteps as shown in the previous example to obtain the gauge interaction terms. First we shall write down the action of the theory which is \cite{Dai:2022whp}
\begin{eqnarray}\label{HDaction}
S_{\mathrm{HD}}&=&\int c~dt\int d^3x~\left[\partial_{\mu}\phi~\partial^{\mu}\phi^{\star}-\frac{1}{\lambda}\left\{\left(\Box\phi\right)\left(\Box\phi\right)^{\star}\right\}-m^2\phi^{\star}\phi\right]\nonumber\\
&=&\int c~dt\int d^3x~\left[\partial_{\mu}\phi~\partial^{\mu}\phi^{\star}-\frac{1}{\lambda}\left\{\left(\partial^{\alpha}\partial_{\alpha}\phi\right)(\partial^{\beta}\partial_{\beta}\phi)^{\star}\right\}-m^2\phi^{\star}\phi\right]~.
\end{eqnarray}
In the above, $\lambda$ represents the coupling of the higher derivative term. The above action is invariant under the following global $U(1)$ transformations 
\begin{eqnarray}
\phi\rightarrow e^{i\eta}\phi;~\phi^{\star}\rightarrow e^{-i\eta}\phi^{\star}~.
\end{eqnarray} 
Before we proceed to the iterative approach, we shall derive the form of the Noether current for a higher derivative theory such as above.
Let us consider a theory which can be generally represented by the following action
\begin{eqnarray}
S[\phi,\phi^{\star}]=\int d^dx~\mathcal{L}\left(\phi,\phi^{\star},\partial_{\mu}\phi,\partial_{\mu}\phi^{\star},\partial_{\mu}\partial_{\nu}\phi,\partial_{\mu}\partial_{\nu}\phi^{\star}\right)\nonumber\\
\end{eqnarray}
Now if we consider the variation of the action with respect to the fields, the following can be written
\begin{eqnarray}
\delta S[\phi,\phi^{\star}]&=&\int d^dx \left[\frac{\partial \mathcal{L}}{\partial \phi}\delta\phi+\frac{\partial \mathcal{L}}{\partial\left(\partial_{\mu}\phi\right)}\delta\left(\partial_{\mu}\phi\right)+\frac{\partial \mathcal{L}}{\partial\left(\partial_{\mu}\partial_{\nu}\phi\right)}\delta\left(\partial_{\mu}\partial_{\nu}\phi\right)+c.c.\right]\nonumber\\
&=&\int d^dx\left[\frac{\partial \mathcal{L}}{\partial \phi}-\partial_{\mu}\left(\frac{\partial \mathcal{L}}{\partial\left(\partial_{\mu}\phi\right)}\right)+\partial_{\mu}\partial_{\nu}\left(\frac{\partial \mathcal{L}}{\partial\left(\partial_{\mu}\partial_{\nu}\phi\right)}\right)+c.c.\right]\delta\phi\nonumber\\
&+&\int d^dx ~\partial_{\mu}\left[\frac{\partial \mathcal{L}}{\partial\left(\partial_{\mu}\phi\right)}\delta\phi+\frac{\partial \mathcal{L}}{\partial\left(\partial_{\mu}\partial_{\nu}\phi\right)}\partial_{\nu}\left(\delta\phi\right)-\partial_{\nu}\left(\frac{\partial \mathcal{L}}{\partial\left(\partial_{\mu}\partial_{\nu}\phi\right)}\right)\delta\phi+c.c.\right]~.
\end{eqnarray}
For $\delta S[\phi,\phi^{\star}]=0$, the above equation leads us to the following form of the conserved current $j^{\mu}$.
\begin{eqnarray}\label{HDNoether}
\eta j^{\mu}=\frac{\partial \mathcal{L}}{\partial\left(\partial_{\mu}\phi\right)}\delta\phi+\frac{\partial \mathcal{L}}{\partial\left(\partial_{\mu}\partial_{\nu}\phi\right)}\partial_{\nu}\left(\delta\phi\right)-\partial_{\nu}\left(\frac{\partial \mathcal{L}}{\partial\left(\partial_{\mu}\partial_{\nu}\phi\right)}\right)\delta\phi+c.c.
\end{eqnarray}
\end{widetext}
One can also note down the equation of motion for the lagrangian density $\mathcal{L}$ 
\begin{eqnarray}
\frac{\partial \mathcal{L}}{\partial \phi}-\partial_{\mu}\left(\frac{\partial \mathcal{L}}{\partial\left(\partial_{\mu}\phi\right)}\right)+\partial_{\mu}\partial_{\nu}\left(\frac{\partial \mathcal{L}}{\partial\left(\partial_{\mu}\partial_{\nu}\phi\right)}\right)+c.c.=0~.\nonumber\\
\end{eqnarray}
We shall now make use of the conserved current for the iterative process to obtain the gauge interaction terms. It is to be mentioned that unlike the previous example, here we shall perform the computations without separating the time and space components. This we have done for the sake of analytical simplicity. From the action given in eq.\eqref{HDaction}, we first write down the original form of the lagrangian density
\begin{eqnarray}
	\mathcal{L}^{(0)}=\partial_{\mu}\phi~\partial^{\mu}\phi^{\star}-\frac{1}{\lambda}\left\{\left(\partial^{\alpha}\partial_{\alpha}\phi\right)(\partial^{\beta}\partial_{\beta}\phi)^{\star}\right\}-m^2\phi^{\star}\phi~.\nonumber\\
\end{eqnarray}
By following the similar approach as shown before, one obtains the conserved current corresponding to the above lagrangian density and after some algebra this in turn will lead us to the following first correction term
\begin{widetext}
\begin{eqnarray}
\mathcal{L}^{(1)}&=&i\left[\phi\left(\partial^{\mu}\phi\right)^{\star}-\phi^{\star}\left(\partial^{\mu}\phi\right)\right]\mathcal{A}_{\mu}+\frac{i}{\lambda}\Big[2\left(\Box\phi\right)\mathcal{A}_{\mu}\left(\partial^{\mu}\phi\right)^{\star}-2\left(\Box\phi\right)^{\star}\mathcal{A}_{\mu}\left(\partial^{\mu}\phi\right)+\phi^{\star}\left(\partial^{\mu}\mathcal{A}_{\mu}\right)\left(\Box\phi\right)-\phi\left(\partial^{\mu}\mathcal{A}_{\mu}\right)\left(\Box\phi\right)^{\star}\Big]~.\nonumber\\
\end{eqnarray}
As the above obtained form of correction has derivative, we repeat the whole process again until we end up with a term which does not have a derivative. We observe that this process leads us to three more correction terms of the following forms
\begin{eqnarray}
\mathcal{L}^{(2)}&=&\phi^{\star}\phi \mathcal{A}_{\mu}\mathcal{A}^{\mu}+\frac{1}{\lambda}\Big[\phi \mathcal{A}_{\mu}\mathcal{A}^{\mu} \left(\Box\phi\right)^{\star}-2\mathcal{A}_{\mu}\left(\partial^{\mu}\phi\right)\mathcal{A}_{\nu}\left(\partial^{\nu}\phi\right)^{\star}-2\mathcal{A}_{\nu}\left(\partial^{\mu}\phi\right)\mathcal{A}_{\mu}\left(\partial^{\nu}\phi\right)^{\star}\nonumber\\
&&-2\phi^{\star}\left(\partial^{\mu}\mathcal{A}_{\mu}\right)\mathcal{A}_{\nu}\left(\partial^{\nu}\phi\right)-2\phi\left(\partial^{\mu}\mathcal{A}_{\mu}\right)\mathcal{A}_{\nu}\left(\partial^{\nu}\phi\right)^{\star}+\phi^{\star} \mathcal{A}_{\mu}\mathcal{A}^{\mu}\left(\Box\phi\right)\nonumber\\
&&-2\left(\partial^{\mu}\mathcal{A}_{\mu}\right)\left(\partial^{\nu}\mathcal{A}_{\nu}\right)\phi^{\star}\phi\Big]\\
\mathcal{L}^{(3)}&=&\frac{i}{\lambda}\Big[2\phi^{\star}\mathcal{A}_{\mu}\mathcal{A}^{\mu}\mathcal{A}_{\nu}\left(\partial^{\nu}\phi\right)-2\phi \mathcal{A}_{\mu}\mathcal{A}^{\mu}\mathcal{A}_{\nu}\left(\partial^{\nu}\phi\right)^{\star}+2\phi^{\star}\phi\left(\partial^{\nu}\mathcal{A}_{\nu}\right)\mathcal{A}_{\mu}\mathcal{A}^{\mu}-2\phi^{\star}\phi\left(\partial^{\nu}\mathcal{A}_{\nu}\right)\mathcal{A}_{\mu}\mathcal{A}^{\mu}\Big]~~\\
\mathcal{L}^{(4)}&=&-\frac{1}{\lambda}\phi^{\star}\phi \mathcal{A}_{\mu}\mathcal{A}^{\mu}\mathcal{A}_{\nu}\mathcal{A}^{\nu}~.
\end{eqnarray}
\end{widetext}
Next we sum up all of the obtained correction terms along with the original lagrangian density and obtain the following result
\begin{eqnarray}\label{TotalLHD}
	\mathcal{L}_{\mathrm{T}}&=&\mathcal{L}^{(0)}+\mathcal{L}^{(1)}+\mathcal{L}^{(2)}+\mathcal{L}^{(3)}+\mathcal{L}^{(4)}\nonumber\\
	&=&\mathcal{D}_{\mu}\phi~\mathcal{D}^{\mu}\phi^{\star}-\frac{1}{\lambda}\left\{\left(\mathcal{D}^{\alpha}\mathcal{D}_{\alpha}\phi\right)(\mathcal{D}^{\beta}\mathcal{D}_{\beta}\phi)^{\star}\right\}-m^2\phi^{\star}\phi\nonumber\\
\end{eqnarray}
where we have defined $\mathcal{D}_{\mu}=\partial_{\mu}+i\mathcal{A}_{\mu}$ to obtain the minimally coupled form.
\begin{widetext}
We can write down the matter action as
\begin{eqnarray}
S_{\mathrm{matter}}^{\mathrm{HD}}=\int c~dt\int d^3x\left[\mathcal{D}_{\mu}\phi~\mathcal{D}^{\mu}\phi^{\star}-\frac{1}{\lambda}\left\{\left(\mathcal{D}^{\alpha}\mathcal{D}_{\alpha}\phi\right)(\mathcal{D}^{\beta}\mathcal{D}_{\beta}\phi)^{\star}\right\}-m^2\phi^{\star}\phi\right]~.
\end{eqnarray}	
It is to be noted that the above lagrangian density is invariant under the following local gauge transformations
\begin{eqnarray}
\phi\rightarrow e^{i\eta(x^{\mu})}\phi,~\phi^{\star}\rightarrow e^{-i\eta(x^{\mu})}\phi^{\star},~\mathcal{A}_{\mu}\rightarrow \mathcal{A}_{\mu}-\partial_{\mu}\eta(x^{\nu})~.\nonumber\\
\end{eqnarray}
With the higher derivative matter action in hand, one can write down the following action corresponding to SQED with higher derivatives (let us call it HD-SQED action)
\begin{eqnarray}\label{HDSQED}
S_{\mathrm{SQED}}^{\mathrm{HD}}&=&S_{\mathrm{matter}}^{\mathrm{HD}}+S_{\mathrm{gauge}}\nonumber\\
&=&\int c~dt\int d^3x\left[\mathcal{D}_{\mu}\phi~\mathcal{D}^{\mu}\phi^{\star}-\frac{1}{\lambda}\left\{\left(\mathcal{D}^{\alpha}\mathcal{D}_{\alpha}\phi\right)(\mathcal{D}^{\beta}\mathcal{D}_{\beta}\phi)^{\star}\right\}-m^2\phi^{\star}\phi-\frac{1}{4}F_{\mu\nu}F^{\mu\nu}\right]~.
\end{eqnarray}
Now let us proceed to perform the non-relativistic reduction of the above action. At first we perform the non-relativistic reduction of the matter sector and after that we move on to the gauge sector. 
\subsection{Non-relativistic reduction of the theory}
For the sake of computational simplicity, we shall write down above action in the following way
\begin{eqnarray}
S_{\mathrm{matter}}^{\mathrm{HD}}&=&\int c~dt\int d^3x\Bigg[\mathcal{D}_{0}\phi~\mathcal{D}_{0}\phi^{\star}-\mathcal{D}_{i}\phi~\mathcal{D}_{i}\phi^{\star}-\frac{1}{\lambda}\Big\{(\mathcal{D}_{0}\mathcal{D}_{0}\phi)\left(\mathcal{D}_{0}\mathcal{D}_{0}\phi\right)^{\star}-\left(\mathcal{D}_{j}\mathcal{D}_{j}\phi\right)\left(\mathcal{D}_{0}\mathcal{D}_{0}\phi\right)^{\star}\nonumber\\
&&-\left(\mathcal{D}_{0}\mathcal{D}_{0}\phi\right)\left(\mathcal{D}_{k}\mathcal{D}_{k}\phi\right)^{\star}+\left(\mathcal{D}_{j}\mathcal{D}_{j}\phi\right)\left(\mathcal{D}_{k}\mathcal{D}_{k}\phi\right)^{\star}\Big\}-m^2\phi^{\star}\phi\Bigg]~.
\end{eqnarray} 
We now make use of the non-relativistic maps given in eq.\eqref{NRmap} and consider the $c\rightarrow\infty$ limit. This in turn yields the non-relativistic matter sector of the theory
\begin{eqnarray}
S_{\mathrm{matter}}^{\mathrm{HD}}|_{\mathrm{NR}}&=&\int dt\int d^3x \Bigg[\left(\frac{i}{2}\right)\left(1-\frac{m^2}{\lambda}\right)\psi^{\star}\overleftrightarrow{D_t}\psi-\frac{1}{2m}(D_i\psi)(D_i\psi)^{\star}-\frac{1}{2\lambda m}\left(D_j~D_j\psi\right)\left(D_k~D_k\psi\right)^{\star}\Bigg]\nonumber~.\\
\end{eqnarray}
\end{widetext}
In obtaining the above, we have defined $D_{X}=\partial_X+iA_X$. Next, we perform the non-relativistic reduction of the gauge sector by following the same procedure as we have shown before. First, we consider the covariant maps of the electric limit (given in eq.\eqref{Elimit}) and also take the limit $c\rightarrow\infty$.
\begin{widetext}
 This yields the following action in the electric limit
\begin{eqnarray}
S_{\mathrm{SQED}}^{\mathrm{HD}}|_{\mathrm{NR}}=\int dt\int d^3x \Bigg[\left(\frac{i}{2}\right)\left(1-\frac{m^2}{\lambda}\right)\psi^{\star}\overleftrightarrow{\partial_t}\psi-\frac{1}{2m}(\Lambda_i\psi)(\Lambda_i\psi)^{\star}&&-\frac{1}{2\lambda m}\left(\Lambda_j~\Lambda_j\psi\right)\left(\Lambda_k~\Lambda_k\psi\right)^{\star}\nonumber\\
&&+\frac{1}{2}\partial^ia^0\left(\partial_ta_i-\partial_ia_0\right)-\frac{1}{4}f_{ij}f^{ij}\Bigg]\nonumber\\
\end{eqnarray}
where we have also included the Maxwell lagrangian density for the electric limit (eq.\eqref{MaxwellE}).
\end{widetext}
The above action is the Galilean relativistic version of the HD-SQED action (given in eq.\eqref{HDSQED}) as we have performed non-relativistic reduction in both gauge and matter scetor of the theory. On the other hand, by making use of the electric limit relations given in eq.\eqref{ElimitRelations}, one can obtain the same action considering the contravariant maps of the electric limit.
\begin{widetext}
Next, we consider the magnetic limit of the gauge sector. Similar to the example of ordinary $\mathbb{C}$-scalar field, we first explore the contravariant maps. This in turn leads to the following form for the action
\begin{eqnarray}
S_{\mathrm{SQED}}^{\mathrm{HD}}|_{\mathrm{NR}}=\int dt\int d^3x \Bigg[\left(\frac{i}{2}\right)\left(1-\frac{m^2}{\lambda}\right)\psi^{\star}\overleftrightarrow{\partial_t}\psi-\frac{1}{2m}(\tilde{\Lambda}_i\psi)(\tilde{\Lambda}_i\psi)^{\star}&&-\frac{1}{2\lambda m}\left(\tilde{\Lambda}_j~\tilde{\Lambda}_j\psi\right)\left(\tilde{\Lambda}_k~\tilde{\Lambda}_k\psi\right)^{\star}\nonumber\\
&&+\frac{1}{2}\partial_ia_0\left(\partial_ta^i-\partial^ia^0\right)-\frac{1}{4}f_{ij}f^{ij}\Bigg]
\end{eqnarray}	
\end{widetext}
where we have included the modified Maxwell lagrangian density for the magnetic limit which is given in eq.\eqref{MaxwellM}. Further, the same action can also be obtained for the covariant magnetic maps where one has to make use of the magnetic limit relations given in eq.\eqref{MlimitRelations}.
\section{Dirac field theory}\label{Sec3}
The action corresponding to the Dirac field theory has the following form
\begin{eqnarray}
S_{\mathrm{F}}=\int c~dt\int d^3x~ \bar{\psi}\left[i\gamma^{\mu}\partial_{\mu}-m\right]\psi
\end{eqnarray}
where $\psi$ is the four-component spinor field and $\bar{\psi}=\gamma^0\psi^{\dagger}$ is the Dirac adjoint. Further, the $\gamma^{\mu}$ matrices satisfy the Clifford algebra $\{\gamma^{\mu},\gamma^{\nu}\}=2\eta^{\mu\nu}$. The spinor fields $\psi$ and $\bar{\psi}$ have the following global $U(1)$ symmetry
\begin{eqnarray}
\psi\rightarrow e^{i\eta}\psi;~\psi^{\dagger}\rightarrow e^{-i\eta}\psi^{\dagger}~.
\end{eqnarray}
The form of the conserved current invoked by the above global symmetry has the following form
\begin{eqnarray}\label{NoetherDirac}
\eta j^{\mu} = \frac{\partial \mathcal{L}}{\partial\left(\partial_{\mu}\psi\right)}\delta\psi+\frac{\partial \mathcal{L}}{\partial\left(\partial_{\mu}\bar{\psi}\right)}\delta\bar{\psi}~.
\end{eqnarray}
Similar to the previous examples, we now proceed to introduce the gauge interactions by following the iterative approach. We start with the given lagrangian density of the Dirac field theory
\begin{eqnarray}
	\mathcal{L}^{(0)}=i\bar{\psi}\gamma^{\mu}\partial_{\mu}\psi -m\bar{\psi}\psi~.
\end{eqnarray}
By using the above lagrangian density in the form of the Noether current (given in eq.\eqref{NoetherDirac}), we obtain the following current and the correction term $\mathcal{L}^{(1)}$
\begin{eqnarray}
	j^{\mu}=-\bar{\psi}\gamma^{\mu}\psi;~\mathcal{L}^{(1)}=-\bar{\psi}\gamma^{\mu}\psi \mathcal{A}_{\mu}~.
\end{eqnarray}
As it can be seen, the first correction to the given lagrangian density has no derivative. This implies the iteractive process to obtain gauge interactions stops here and we end up with the following action
\begin{eqnarray}\label{OnlyF}
	S_{\mathrm{matter}}^{\mathrm{F}}&=&\int c~dt\int d^3x~\left[\mathcal{L}^{(0)}+\mathcal{L}^{(1)}\right]\nonumber\\
	&=&\int c~dt\int d^3x~\left[i\bar{\psi}\gamma^{\mu}\partial_{\mu}\psi-\bar{\psi}m\psi-\bar{\psi}\gamma^{\mu}\psi\mathcal{A}_{\mu}\right]\nonumber\\
	&=&\int c~dt\int d^3x~\bar{\psi}\left[i\gamma^{\mu}\mathcal{D}_{\mu}-m\right]\psi;~\mathcal{D}_{\mu}=\partial_{\mu}+i\mathcal{A}_{\mu}~.\nonumber\\
\end{eqnarray}
Once again it can be observed that the above action is invariant under the following local gauge transformations
\begin{eqnarray}
\phi\rightarrow e^{i\eta(x^{\mu})}\phi;~\phi^{\star}\rightarrow e^{-i\eta(x^{\mu})}\phi^{\star};~\mathcal{A}_{\mu}\rightarrow \mathcal{A}_{\mu}-\partial_{\mu}\eta(x^{\nu})~.\nonumber\\
\end{eqnarray}
where $\mathcal{A}_{\mu}=(\mathcal{A}_0,\mathcal{A}_i)\equiv\left(\frac{A_0}{c},A_i\right)$. Further, if we take into account the dynamical term for the $U(1)$ gauge field $\mathcal{A}_{\mu}$, we obtain the QED action. This reads
\begin{eqnarray}\label{QED}
S_{\mathrm{QED}}=\int c~dt\int d^3x~\left[i\bar{\psi}\gamma^{\mu}\mathcal{D}_{\mu}-\bar{\psi}m\psi-\frac{1}{4}F_{\mu\nu}F^{\mu\nu}\right]~.\nonumber\\
\end{eqnarray}
We now proceed to non-relativistic reduction of both matter and gauge sectors of the above given form of QED action.
\subsection{Non-relativistic reduction}
Firstly, we focus on the matter sector of the QED action. For the sake of computational simplicity, we first recast matter largrangian density to the following form
\begin{eqnarray}\label{FmatterAction}
	\mathcal{L}_{\mathrm{matter}}^{\mathrm{F}}&=&i\bar{\psi}\gamma^{\mu}\partial_{\mu}\psi-\bar{\psi}m\psi-\bar{\psi}\gamma^{\mu}\psi\mathcal{A}_{\mu}\nonumber\\
	&=&ic\psi^{\dagger}\mathcal{D}_0\psi+ic\psi^{\dagger}\alpha_i\mathcal{D}_i\psi-mc^2\psi^{\dagger}\beta\psi~.
\end{eqnarray}
In obtaining the above form, we have used the following identities
\begin{eqnarray}
\left(\gamma^0\right)^2=\begin{pmatrix}
\mathbb{1} & 0\\
0 & \mathbb{1}
\end{pmatrix};~	
\alpha_i\equiv\gamma^0\gamma^i=\begin{pmatrix}
0 & \sigma_i\\
\sigma_i & 0
\end{pmatrix};\nonumber\\
\beta\equiv\gamma^0=\begin{pmatrix}
\mathbb{1} & 0\\
0 & -\mathbb{1}
\end{pmatrix}~.
\end{eqnarray}
Further, we also restored the speed of light $c$.\\
We now write down the four-component Dirac spinor $\psi$ into two Weyl spinors $\xi$ and $\chi$ as
\begin{eqnarray}\label{FWmap}
\psi = \begin{pmatrix}
\xi\\
\chi
\end{pmatrix}~.	
\end{eqnarray}
\begin{widetext}
By substituting the above into the matter action (given in eq.\eqref{FmatterAction}), we obtain the following form
\begin{eqnarray}\label{LMatter}
\mathcal{L}_{\mathrm{matter}}^{\mathrm{F}}&=&\begin{pmatrix}
\xi^{\dagger}&\chi^{\dagger}
\end{pmatrix}
\left[ic\begin{pmatrix}
\mathcal{D}_0\xi\\
\mathcal{D}_0\chi
\end{pmatrix}+ic\begin{pmatrix}
0 & \sigma_i\\
\sigma_i & 0
\end{pmatrix}
\begin{pmatrix}
\mathcal{D}_i\xi\\
\mathcal{D}_i\chi
\end{pmatrix}-mc^2\begin{pmatrix}
\mathbb{1} & 0\\
0 & \mathbb{1}
\end{pmatrix}
\begin{pmatrix}
\xi\\
\chi
\end{pmatrix} \right]\nonumber\\
&=& ic\left[\left\{\xi^{\dagger}\mathcal{D}_0\xi+\chi^{\dagger}\mathcal{D}_0\chi\right\}+\left\{\xi^{\dagger}\sigma_i\mathcal{D}_i\chi+\chi^{\dagger}\sigma_i\mathcal{D}_i\xi\right\}\right]-mc^2\left[\xi^{\dagger}\xi-\chi^{\dagger}\chi\right]~.
\end{eqnarray}
\end{widetext}
To proceed further, we now introduce the following non-relativistic maps \cite{Banerjee:2019hec}
\begin{eqnarray}\label{NRmaps}
\xi=\frac{1}{\sqrt{2c}}e^{-imc^2t}\xi_0(t,x^i);~\chi=\frac{1}{\sqrt{2c}}e^{-imc^2t}\chi_0(t,x^i)~.\nonumber\\
\end{eqnarray}
Furthermore, we also make use of the following identity
 \begin{eqnarray}
\mathcal{D}_0&=& \partial_0+i\mathcal{A}_{\mu}=\left(\frac{1}{c}\right)\left[\partial_t+iA_0\right]\equiv\left(\frac{1}{c}\right)D_t ~.
\end{eqnarray}
\begin{widetext}
By incorporating these in the lagrangian density (eq.\eqref{LMatter}) we obatin the following result
\begin{eqnarray}\label{LFM}
\mathcal{L}_{\mathrm{matter}}^{\mathrm{F}}|_{\mathrm{NR}}&=&\left(\frac{i}{2c}\right)\left[-imc^2\xi_0^{\dagger}\xi_0+\xi_0^{\dagger}D_t\xi_0-imc^2\chi_0^{\dagger}\chi_0+\chi_0^{\dagger}D_t\chi_0\right]\nonumber\\
&&+\left(\frac{ic}{2c}\right)\left[\xi_0^{\dagger}\sigma_iD_i\chi_0+\chi_0^{\dagger}\sigma_iD_i\xi_0\right]	-\frac{mc^2}{2c}\left[\xi_0^{\dagger}\xi_0-\chi_0^{\dagger}\chi_0\right]\nonumber\\
&=&\left(\frac{i}{2c}\right)\left[\xi_0^{\dagger}D_t\xi_0+\chi_0^{\dagger}D_t\chi_0\right]+\left(\frac{i}{2}\right)\left[\xi_0^{\dagger}\sigma_iD_i\chi_0+\chi_0^{\dagger}\sigma_iD_i\xi_0\right]+\textcolor{blue}{mc\chi_0^{\dagger}\chi_0}~.
\end{eqnarray}
\end{widetext}
It can be observed that in the above lagrangian density the two-component spinors are in mixed form. However, we want to separate them and in order to do that we have to use the on-shell condition, more precisely, we have to make use of the equation of motion (EoM) associated to the lagrangian density $\mathcal{L}_{\mathrm{matter}}^{\mathrm{F}}$. This can be obtained by considering variation of the action given in eq.\eqref{QED} with respect to the spinor field $\psi$ (or $\bar{\psi}$). This yields the following Dirac equation for the QED lagrangian density
\begin{eqnarray}
&&\left[ic\gamma^{\mu}\mathcal{D}_{\mu}-mc^2\right]\psi=0\nonumber\\
\Rightarrow&& iD_t\psi+ic\vec{\alpha}.\vec{D}\psi=\beta mc^2\psi~.
\end{eqnarray}
\begin{widetext}
We now substitute the Foldy-Wouthusyen relation (given in eq.\eqref{FWmap}) in the Dirac equation along with the non-relativistic maps (given in eq.\eqref{NRmaps}). This in turn yields
\begin{eqnarray}
\begin{pmatrix}
iD_t\xi_0+mc^2\xi_0\\
iD_t\chi_0+mc^2\chi_0
\end{pmatrix}+ic
\begin{pmatrix}
0 & \sigma_i\\
\sigma_i & 0
\end{pmatrix}
\begin{pmatrix}
D_i\xi_0\\
D_i\chi_0
\end{pmatrix}-mc^2
\begin{pmatrix}
\xi_0\\
-\chi_0
\end{pmatrix}=0~.\nonumber\\
\end{eqnarray}	
\end{widetext}
It is to be noted that one can write two coupled algebraic equations from the above form of the Dirac equation. This reads
\begin{eqnarray}
	iD_t\xi_0+ic\sigma_iD_i\eta_0=0\label{eq1}\\
	iD_t\chi_0+ic\sigma_iD_i\xi_0+2mc^2\chi_0=0\label{eq2}~.
\end{eqnarray}
The second equation can be recast to the following form
\begin{eqnarray}
	\chi_0=-\frac{i}{2mc^2}D_t\chi_0-\frac{i}{2mc}\sigma_iD_i\xi_0~.
\end{eqnarray}
This in turn means in the $c\rightarrow\infty$ limit, one can isolate the leading piece and write down the following relation
\begin{eqnarray}\label{PSrelation}
	\chi_0\approx -\frac{i}{2mc}\sigma_iD_i\xi_0~.
\end{eqnarray}
We now substitute this relation in eq.\eqref{eq1} and obtain the following spin modified Schr\"odinger equation which was first proposed by Pauli in \cite{Pauli1927ZurQD}
\begin{eqnarray}\label{PSeq}
iD_t\xi_0&&=-\frac{1}{2m}\left(\sigma_i\sigma_j\right)\left(D_iD_j\xi_0\right)\nonumber\\
&&=-\frac{1}{2m} \left(\vec{D}^2+\frac{i}{2}\epsilon_{ijk}\sigma_k F_{ij}\right)\xi_0~.
\end{eqnarray}
This equation is generally denoted as the Pauli equation or Pauli-Schr\"odinger equation \cite{Cohen-Tannoudji1977Quantum}. It was introduced in order to describe the intrinsic magnetic moment of the electron by using the standard non-relativistic framework.
\begin{widetext}
Next, we substitute the relation that we have obtained in eq.\eqref{PSrelation} in the Lagrangian density given in eq.\eqref{LFM}
\begin{eqnarray}
\mathcal{L}_{\mathrm{matter}}^{\mathrm{F}}|_{\mathrm{NR}}&=&\left(\frac{i}{2c}\right)\Big[\xi_0^{\dagger}D_t\xi_0+\left(\frac{1}{2mc}\right)^2\left(D_i\xi\right)^{\dagger}\sigma_i^{\dagger}\sigma_jD_jD_t\xi_0-\left(\frac{i}{2m}\right)\xi_0^{\dagger}\left(\sigma_i\sigma_j\right)\left(D_iD_j\xi_0\right)\nonumber\\
&&+\left(\frac{i}{2m}\right)\left(D_i\xi_0\right)^{\dagger}\sigma_i^{\dagger}\sigma_jD_j\xi_0-\left(\frac{i}{2m}\right)\left(D_i\xi_0\right)^{\dagger}\sigma_i^{\dagger}\sigma_jD_j\xi_0\Big]~.
\end{eqnarray}
We now substitute the above form of the lagrangian in the action (given in eq.\eqref{OnlyF}) and consider the $c\rightarrow\infty$ limit. This results in 
\begin{eqnarray}
S_{\mathrm{matter}}^{\mathrm{F}}|_{\mathrm{NR}}&=&\int dt\int d^3x~\frac{i}{2}\Big[\xi_0^{\dagger}D_t\xi_0-\left(\frac{i}{2m}\right)\xi_0^{\dagger}\left(\sigma_i\sigma_j\right)\left(D_iD_j\xi_0\right)\Big]~.\nonumber\\
\end{eqnarray}
Let us now focus on the last term of the above action as it can be simplified in the following way
\begin{eqnarray}
	&&\left(\frac{i}{2m}\right)\xi_0^{\dagger}\left(\sigma_i\sigma_j\right)\left(D_iD_j\xi_0\right)\nonumber\\
	=&&\left(\frac{i}{2m}\right)\xi_0^{\dagger}\left(\delta_{ij}\mathbb{1}+i\epsilon_{ijk}\sigma_k\right)\left(D_iD_j\xi_0\right)\nonumber\\
	=&&\left(\frac{i}{2m}\right)\xi_0^{\dagger}\left(\vec{D}^2+\frac{i}{2}\epsilon_{ijk}\sigma_k\left[D_i,D_j\right]\right)\xi_0\nonumber\\
	=&&\left(\frac{i}{2m}\right)\xi_0^{\dagger}\left(\vec{D}^2+\frac{i}{2}\epsilon_{ijk}\sigma_k F_{ij}\right)\xi_0~.
\end{eqnarray}
By incorporating the above simplification in the action $S_{\mathrm{matter}}^{\mathrm{F}}|_{\mathrm{NR}}$, we obtain the following expression
\begin{eqnarray}
S_{\mathrm{matter}}^{\mathrm{F}}|_{\mathrm{NR}}&=& \int dt\int d^3x~\left(\frac{i}{2}\right)\Big[\xi_0^{\dagger}D_t\xi_0-\left(\frac{i}{2m}\right)\xi_0^{\dagger}\left(\vec{D}^2+\frac{i}{2}\epsilon_{ijk}\sigma_k F_{ij}\right)\xi_0\Big]~.
\end{eqnarray}
This is the final version of the non-relativistic matter action. This is nothing but the Pauli-Schr\"odinger action. Our next task is to perform non-relativistic reduction of the gauge sector. Similar to the previous examples, we first consider the covariant maps of the electric limit.
By using the non-relativistic maps of the gauge field (given in eq.\eqref{Elimit}) and considering the limit $c\rightarrow\infty$, we obtain the following Galilean relativistic QED action
\begin{eqnarray}\label{GPS}
S_{\mathrm{QED}}^{\mathrm{NR}}=\int dt\int d^3x~\left[\frac{i}{2}\xi_0^{\dagger}\partial_t\xi_0+\left(\frac{1}{4m}\right)\xi_0^{\dagger}\left(\vec{\Lambda}^2+\frac{i}{2}\epsilon_{ijk}\sigma_k f_{ij}\right)\xi_0+\frac{1}{2}\partial^ia^0\left(\partial_ta_i-\partial_ia_0\right)-\frac{1}{4}f_{ij}f^{ij}\right]~.
\end{eqnarray}
\end{widetext}
The above action can be denoted as the Galilean Pauli-Schr\"odinger action (in the electric limit) as in the usual Pauli-Schr\"odinger action only the matter part is non-relativistic but the gauge part (or electromagnetic part) is the standard relativistic electromagnetism. Here, we have performed non-relativistic reduction of both sectors (matter and gauge) of the standard QED theory and this leads us to a action which is invariant under Galilean relativity. Furthermore, the EoM corresponding to the spinor component $\xi_0$ of the above action can be obtained by taking the electric limit of standard Pauli equation (given in eq.\eqref{PSeq}). This reads
\begin{eqnarray}\label{GPeq}
	i\partial_t\xi_0&=&-\frac{1}{2m}\left(\vec{\Lambda}^2+\frac{i}{2}\epsilon_{ijk}\sigma_k f_{ij}\right)\xi_0~.
\end{eqnarray}
This EoM can be denoted as the Galilean Pauli equation (in the electric limit) as the background gauge fields ($a_i$) is now non-relativistic. It is also to be noted that similar to the previous examples, one can also obtain the above action (and also the EoM given in eq.\eqref{GPeq}) by considering the caontravariant maps of the electric limit. In order to do that one has to make use of the relations given in eq.\eqref{ElimitRelations}.
\begin{widetext}
 On the other hand, for the magnetic contravariant maps, we get
\begin{eqnarray}\label{GPSm}
S_{\mathrm{QED}}^{\mathrm{NR}}=\int dt\int d^3x~\left[\frac{i}{2}\xi_0^{\dagger}\partial_t\xi_0+\left(\frac{1}{4m}\right)\xi_0^{\dagger}\left(\vec{\tilde{\Lambda}}^2+\frac{i}{2}\epsilon_{ijk}\sigma_k f_{ij}\right)\xi_0+\frac{1}{2}\partial_ia_0\left(\partial_ta^i-\partial^ia^0\right)-\frac{1}{4}f_{ij}f^{ij}\right]~.
\end{eqnarray}	
\end{widetext}
This is the Galilean Pauli-Schr\"odinger action (in the magnetic limit) and corresponding EoM for the spin-component $\xi_0$ reads
\begin{eqnarray}
i\partial_t\xi_0&=&-\frac{1}{2m}\left(\vec{\tilde{\Lambda}}^2+\frac{i}{2}\epsilon_{ijk}\sigma_k f_{ij}\right)\xi_0~.
\end{eqnarray}
As we have already mentioned in the previous examples, the above action can also be obtained from the magnetic convariant maps and for that one needs to use the relations given in eq.\eqref{MlimitRelations}.
\section{Conclusion}\label{Sec4}
We consider a class of well-known relativistic matter actions which have global $U(1)$ phase symmetries, namely, the free complex ($\mathbb{C}$) scalar field action, $\mathbb{C}$-scalar field action with higher derivative corrections and the free Dirac field action. The standard approach to introduce the minimal coupling or more simply, to make the global $U(1)$ symmetry of the free matter theory a local one, one has to rely upon the gauge principle. This principle tells that the covariant derivatives are used in place of ordinary derivatives of the free matter theory and the additional connection terms make up for the variation in the matter field's local transformation. In this work, we follow a different approach, more precisely, an iterative Noether approach to obtain the required minimal couplings which are invariant under local gauge transformations. This method was first shown by Deser in \cite{Deser:1969wk} where one follows a systematic step by step iterative process based upon the the deifition of Noether current.\\
We first apply this method to the usual relativistic free $\mathbb{C}$-scalar field theory and obtain the scalar quantum electrodynamics (SQED) action. In order to exploit the assocaited low energy effective description of the theory, we then introduce the non-relativistic map (NR map) for the matter sector and take the $c\rightarrow\infty$ limit and obtain an action which represents a non-relativistic matter field theory coupled to a relativistic gauge field (standard electromagnetic theory). We observe that the resulting theory still has covariant derivatives in both time and space part, however, the time-part is left with only single derivative while the space part still has double derivatives. This in turn means it is nothing but the Schr\"odinger action coupled to standard electromagnetic fields. Further, we perform non-relativistic reduction for the gauge sector by following the approach shown in \cite{Banerjee:2022eaj}. We exploit both electric and magnetic limits for the electromagnetic gauge theory by making use of the corresponding NR maps and consider once again the $c\rightarrow\infty$. This results in a complete Galilean relativistic SQED (NR-SQED) action which can also be denoted as the Schr\"odinger action coupled to Galilean electromagnetism. Furthermore, we note that in resulting action, the time-covariance is broken which is being represented by the fact that the temporal derivative is an ordinary one while the spatial-covariance is still intact. This observation also supports the fact that in a complete Galilean framework, time shall have an absolute status. On the other hand, in this reduction procedure, we observe that both electric and magnetic limits have different maps for the original relativistic covariant ($A_{\mu}$) and contravariant components ($A^{\mu}$) and by demanding that they shall lead to the same resulting action, we obtain a set of relation which exist between the non-relativistic temporal components ($a_0$ and $a^0$) and between the non-relativistic spatial components ($a_i$ and $a^i$). For the electric limit, we name these relations as the electric limit relations and for the magnetic limit, we call them magnetic limit relations.\\
Next, we move onto consider a free theory with bit more complicated form, namely, relativistic free $\mathbb{C}$-scalar field action with higher derivative (HD) corrections. The iterative Noether approach in this case leads us to the SQED action with HD corrections (HD-SQED). We then follow the same approach for NR reductions that we have shown for the standard SQED theory and end up with the Schr\"odinger action (in the presence of Galilean electromagnetism) with HD corrections. Once again we observe that the resulting action has oridinary time derivative (with the cofficient now modified due to the inclusion of the HD terms) but all of the spatial derivatives are covariant one.\\
Finally, we consider the free Dirac field theory and obtain the standard QED action by iteratively introducing the interaction term. To proceed further, we then use a Foldy-Wouthuysen type expansion in order to map the four-component spinor $\psi$ to the two-component form and consider the $c\rightarrow\infty$ limit. By using the on-shell condition or more precisely the equation of motion (Dirac equation), we obtain the Pauli-Schr\"odinger action. It is very interesting to observe that while the QED action contains single covariant derivatives associated to both time and space, the Pauli-Schr\"odinger action has single time-covariant derivative but double spatial covariant derivative. Similar to the previous examples, we then proceed to perform the NR for the gauge sectors by utilizing the maps corresponding to both electric and magnetic limits. The final result is very subtle as it is a Galilean version of the standard Pauli-Schr\"odinger action which has a ordinary single time derivative, double covariant spatial derivatives and the gauge field is Galilean electromagnetism. The obtained set of low energy Galilean effective actions and their properties are novel to the literature which we believe will be very useful as they represent proper dynamics in the Galilean framework.  
\section{Acknowledgments}
AS would like to thank S.N. Bose National Centre for Basic Sciences for the financial support through its Advanced Postdoctoral Research Programme. The authors would like to thank the anonymous referee for the useful comments.
\section*{Appendix: Gauging the Schr\"odinger theory by Deser's iterative Noether approach}
We first consider the following non-relativistic free matter action
\begin{eqnarray}
	S_{NR}=\int dt\int d^3x\left(\frac{i}{2}\right)\left[\psi^{*}\overleftrightarrow{\partial_t}\psi+\frac{i}{m}(\partial_i\psi)(\partial_i\psi)^{\star}\right]~.\nonumber\\
\end{eqnarray}
Now we proceed to apply Deser's iterative Noether approach to derive the gauge interactions. We start with the time part of the Lagrangian density, that is,
\begin{eqnarray}
	\mathcal{L}_t=\frac{i}{2}\left(\psi^{*}\partial_t\psi-\psi\partial_t\psi^*\right)~.
\end{eqnarray}
The definition of Noether current gives us the following results for current and correction to the Lagrangian density
\begin{eqnarray}
	j^t&=&-i\psi^*\psi\\
	\mathcal{L}_t^{(1)}&=&j^t\mathcal{A}_0\equiv -i\psi^*\psi \mathcal{A}_0~.
\end{eqnarray}
This gives the final form for the time part of the Lagrangian density to be
\begin{eqnarray}
	\mathcal{L}_t+\mathcal{L}_t^{(1)}=\frac{i}{2}\left(\psi^{*}\partial_t\psi-\psi\partial_t\psi^*-2\psi^*\psi \mathcal{A}_0\right)~.
\end{eqnarray}
It is to be noted that the correction $\mathcal{L}_t^{(1)}$ produces a vanishing result when the Noether current formula is applied once again. So, the iteration stops here. We now focus on the space part of the Lagrangian density, that is,
\begin{eqnarray}
	\mathcal{L}_s=-\frac{1}{2m} (\partial_i\psi)(\partial_i\psi)^{\star}~.
\end{eqnarray}
It can be easily shown that one obtains the following correction terms by using the iterative process to $\mathcal{L}_s$
\begin{eqnarray}
	\mathcal{L}_s^{(1)}&=&-\frac{1}{2m}\left[\left(\partial_i\psi\right)^*\psi-\left(\partial_i\psi\right)\psi^*\right]\mathcal{A}_i\\
	\mathcal{L}_s^{(2)}&=&\frac{1}{2m}\left(\psi^*\psi\right)\mathcal{A}_i^2~.
\end{eqnarray}
\begin{widetext}
We now have all of the necessary ingredients to write down the final form of the action corresponding to an interacting Schr\"odinger theory
\begin{eqnarray}
	S_{\mathrm{matter}}^{\mathrm{NR}}&=&\int dt\int d^3x \left(\mathcal{L}_t+\mathcal{L}_t^{(1)}+\mathcal{L}_s+\mathcal{L}_s^{(1)}+\mathcal{L}_s^{(2)}\right)\nonumber\\
	&=& \int dt\int d^3x \left(\frac{i}{2}\right)\left[\psi^{*}\overleftrightarrow{\partial_t}\psi-2\psi^*\psi \mathcal{A}_0+\frac{i}{m}(\partial_i\psi)(\partial_i\psi)^{\star}+\frac{i}{m}\left(\left(\partial_i\psi\right)^*\psi-\left(\partial_i\psi\right)\psi^*\right)\mathcal{A}_i-\frac{i}{m}\left(\psi^*\psi\right)\mathcal{A}_i^2\right]~.\nonumber\\
\end{eqnarray}
Now using the contravariant maps for the magnetic limit and considering the $c\rightarrow\infty$ limit, the above action takes the following form
\begin{eqnarray}
S_{\mathrm{matter}}^{\mathrm{NR}}=\int dt\int d^3x \left(\frac{i}{2}\right)\left[\psi^{\star}\left(\partial_t\psi\right)-\psi\left(\partial_t\psi\right)^{\star}+\frac{i}{m}|\left(\partial_i-ia^{i}\right)\psi|^2\right]~.
\end{eqnarray}
 Finally, the above form of the action can be written down in terms of the gauge invariant minimal couplings in the following way
\begin{eqnarray}\label{gauged}
	S_{\mathrm{matter}}^{\mathrm{NR}}=\int dt\int d^3x \left(\frac{i}{2}\right)\left[\psi^{\star}\overleftrightarrow{\partial_t}\psi+\frac{i}{m}(\tilde{\Lambda}_i\psi)(\tilde{\Lambda}_i\psi)^{\star}\right];~\tilde{\Lambda}_i=\partial_i-ia^i~.
\end{eqnarray}	
\end{widetext}
Eq.\eqref{gauged} is the gauged action given in eq.\eqref{Mcontra} of the main text. Eq.\eqref{magneticcova} in the main text follows from the above action using the magnetic relations between the gauge components (given in eq.\eqref{MlimitRelations}). One can similarly use the electric covariant/contravariant maps together with the relations given in eq.\eqref{ElimitRelations} to obtain the form of the action given in eq.\eqref{Ecovar}.

\bibliographystyle{hephys.bst}
\bibliography{References.bib}
\end{document}